\documentclass[aps,preprint,showpacs,superscriptaddress]{revtex4-1}  
\usepackage{graphicx}
\usepackage{subcaption}
\usepackage{bm}   
\usepackage{color}
\usepackage{amssymb}
\usepackage{amsmath}
\usepackage{float}

\newcommand{\e}{\normalfont\mbox{e}\,}

\begin{document}	
	
	\title{Formation of rogue waves on the periodic background in a fifth-order nonlinear Schr\"odinger equation}
	
\author{N. Sinthuja}
\affiliation{Department of Nonlinear Dynamics, Bharathidasan University, Tiruchirappalli - 620 024, Tamilnadu, India}
\author{K. Manikandan}
\affiliation{Department of Nonlinear Dynamics, Bharathidasan University, Tiruchirappalli - 620 024, Tamilnadu, India}
\author{M. Senthilvelan}
\affiliation{Department of Nonlinear Dynamics, Bharathidasan University, Tiruchirappalli - 620 024, Tamilnadu, India}

\vspace{10pt}
	
\begin{abstract}
We construct rogue wave solutions of a fifth-order nonlinear Schr\"odinger equation on the Jacobian elliptic function background. By combining Darboux transformation and the nonlinearization of spectral problem, we generate rogue wave solution on two different periodic wave backgrounds. We analyze the obtained solutions for different values of system parameter and point out certain novel features of our results.  We also compute instability growth rate of both $dn$ and $cn$ periodic background waves for the considered system through spectral stability problem.  We show that instability growth rate decreases (increases) for $dn$-$(cn)$ periodic waves when we vary the value of the elliptic modulus parameter.
\end{abstract}

\maketitle
\section{Introduction}

During the past two decades, both experimentalists and theoreticians have been inclined to investigate the rogue wave (RW) phenomenon on different perspectives because of its importance in their concerned fields \cite{ankiew,obsorne,dudley,cha,so,kib,mani,blu}.   RW is traditionally defined by `` a wave that appears from nowhere and disappears without a trace" \cite{nak}.  The formation of RWs can be related to the modulation instability of the background wave \cite{chen4,ye,zakharov}. Based on the theoretical arguments it has been shown that RWs can arise on constant, multi-soliton and periodic backgrounds \cite{agaf,mu,agaf1,mani2,chen5}. On the other hand, in nature,  the waves that appear on the surface of the ocean may also come-up in a periodic form. To study the emergence of RWs on the periodic wave background, very recently, investigations have been made to construct RWs with a periodic wave background. In this direction, initially, the existence of RWs on the periodic background was studied in the nonlinear Schr\"odinger (NLS) equation only  through a numerical scheme \cite{kedziora}. Later, an explicit expression for the RW solution on the periodic background has been derived analytically for the same equation \cite{chen2}. To construct RW solution with a periodic background, the authors of Ref. \cite{chen2} have combined two algebraic techniques, namely the Darboux transformation (DT) technique and the method of nonlinearization of spectral problem. By combining these two approaches, RW solutions on the periodic background have been constructed for the NLS equation, modified Korteweg-de-Vries (KdV) equation, complex modified KdV equation and sine-Gordon equation \cite{chen1,chen3,sinthu3,li}. Subsequently, RWs on the double periodic (both in $x$ and $t$) background has also been derived for the NLS equation \cite{chen5}.

To describe the dynamics of ultrashort pulse propagation in optical fibers and in the study of the nonlinear spin excitations in Heisenberg ferromagnetic spin chain, certain higher-order terms (self-steepening, self-frequency shift, third, fourth and fifth-order dispersions) have been included in the well-known NLS equation, see for example Refs. \cite{hirota, agrawal,akmv1,akmv2,porse,porse1,wzzhao,radha,akmv3,akmv4,sun,song}.  These higher order terms played a crucial role in bringing out certain properties associated with the pulse/wave propagation in the respective problem.  Interestingly, RW solutions have also been constructed for some of these higher-order NLS equations, say for example Hirota equation, a fourth-order NLS equation and Ito equation \cite{peng,zhang,zhagao,sin}.  In the present work, we consider a fifth-order nonlinear Schr\"odinger (FONLS) equation that describes a nonlinear spin excitation in the Heisenberg ferromagnetic spin chain and construct RW solutions on the periodic background.  

Recently, the authors of Ref. \cite{wang} have investigated the higher-order RW solutions of this equation through DT method. The authors have reported RW solutions only on the constant wave background. To the best of our knowledge, RW solutions on the periodic wave background for this FONLS equation  is yet to be reported. Hence, we intend to construct such solutions and study their dynamical features. The results coming out from this study may help to understand some features that are associated with RWs which emerge from higher order effects. 

In this work, we construct RW solution on the $dn$ and $cn$ periodic wave background for the aforementioned  FONLS equation by suitably combining the DT and the method of  nonlinearization of Lax pair \cite{chen2}. We start our analysis by deriving periodic travelling wave solutions of the considered equation.  We then consider the nonlinearization of Lax pair method.  By exploiting the interconnection between the squared eigenfunctions and the potential that appear in the spectral problem we derive certain differential constraints.  By appropriately solving these constraints we identify the eigenvalues and squared eigenfunctions that correspond to the travelling wave solution which we found earlier in terms of Jacobian elliptic functions, namely $dn$ and $cn$. Substituting back the obtained eigenvalues, eigenfunctions and the seed/periodic travelling wave solution in the one-fold DT formula of the FONLS equation, we can create the periodic background. We generate the RWs on top of this periodic wave background by constructing another independent solution to the spectral problem for the same eigenvalues that we determined earlier for the travelling wave solutions.  We analyze the obtained RW solutions for three different values of the system parameter.  The outcome shows that the frequency of the periodic background wave increases in the $(x-t)$ plane when we increase the value of the system parameter. Our results also reveal that the amplitude of RWs on the periodic background diminishes (enhances) for $dn$ ($cn$) waves when we vary the elliptic modulus parameter $(k)$, from a lower value to a higher value. 

Further, we also investigate the modulation instability of the FONLS equation by analyzing the spectral stability problem.  To begin, we solve the Lax pair equations analytically by considering periodic wave solutions and determine the eigenvalues.  In this process we also determine the values of  certain unknown parameters that appear in the solution as well. To understand the stability of the considered background solutions, we linearly perturb the periodic wave (seed) solutions. Using this perturbed solution we construct spectral stability problem and compute numerically the instability rate of both $dn$ and $cn$ periodic waves.  When we increase the $k$ value from $0$ to $1$ the instability growth rate decreases for the $dn$-periodic waves and the instability growth rate increases for the $cn$-periodic waves.  Our results also confirm that the height of RW on the $dn$-periodic background is higher than that of the RW which appear on the $cn$-periodic background. 
 
\par We organize our work as follows: In Sec. 2, we consider the FONLS equation and construct RW solutions on the periodic wave background. We present Lax pair and one-fold Darboux transformation for the considered equation. In Sec. 3, we derive the periodic travelling wave solutions of FONLS equation and determine the associated eigenvalues and squared eigenfunctions through the method of nonlinearization of spectral problem. In Sec.4, we determine the second linearly independent solution to the spectral problem for the same eigenvalue and achieve the desired task. In Sec. 5, we calculate the instability rates for both $dn$- and $cn$- periodic waves of FONLS equation.   We summarize our results in Sec. 6.
\section{Model and one-fold Darboux Transformation}
The nonlinear spin excitations in the Heisenberg ferromagnetic spin chain can be described by the following generalized FONLS equation \cite{wzzhao,radha,sun,song,wang,feng}, namely 
\begin{align}
ir_t+r_{xx}+2|r|^2 r-i(\epsilon(r_{xxxxx}+10|r|^2 r_{xxx}+20 \bar{r} r_x  r_{xx}+30|r|^4 r_x \nonumber\\+10(r|r_x|^2 )_x)+r_x)=0,
\label{e1}
\end{align}
where $r(x,t)$ represents the complex wave envelope, $x$ and $t$ describes the spatial and temporal coordinates, subscripts denote partial derivatives with respect to $x$ and $t$  respectively, and $\epsilon$ is an arbitrary real parameter.  Equation (\ref{e1}) has been obtained by deforming the inhomogeneous Heisenberg ferromagnetic system through the space curve formalism and using the prolongation structure theory.  Equation (\ref{e1}) also describes the propagation of attosecond pulses in an inhomogeneous optical fiber when the duration of optical pulses close to $20 fs$, see for example, Refs. \cite{akmv3,akmv4,backus,christov,henkel,wang2}.

 Equation (\ref{e1}) possesses a $(2\times 2)$ Lax pair of the form \cite{feng}
\begin{subequations}
	\label{refno}
	\begin{align}
	\label{e2}
	\varphi_{x}=U(\lambda,r) \varphi,\qquad U(\lambda,r)=\begin{pmatrix} \lambda & r\\ -\bar{r} & -\lambda \end{pmatrix}, 
	\end{align}
	\begin{align}
	\label{e3} 
	\varphi_{t}=V(\lambda,r) \varphi,\qquad V(\lambda,r)=\begin{pmatrix} A & B\\ C & -A \end{pmatrix}, 
	\end{align}
	\text{with}
	\begin{align}
	\label{e4} 
	A=&16\lambda^5\epsilon+8\lambda^3\epsilon|r|^2-4\lambda^2\epsilon(r\bar{r}_x-r_x \bar{r})+2i\lambda^2+2\lambda\epsilon(r\bar{r}_{xx}+\bar{r}r_{xx}-|r_x|^2+3|r|^4)\nonumber \\
	&+\lambda+\epsilon(\bar{r}r_{xxx}-r\bar{r}_{xxx}+r_x \bar{r}_{xx}-\bar{r}_x r_{xx}+6|r|^2\bar{r}r_x-6|r|^2r \bar{r}_x)+i|r|^2, \\
	B=&16\lambda^4\epsilon r+8\lambda^3\epsilon r_x+4\lambda^2\epsilon(r_{xx}+2|r|^2r)+2\lambda\epsilon(r_{xxx}+6|r|^2r_x)+2i\lambda r\nonumber\\&+\epsilon(r_{xxxx}+8|r|^2r_{xx}+2r^2\bar{r}_{xx}+4|r_x|^2r+6{r}^2_x\bar{r}+6|r|^4r)+ir_x+r, \\ 
	C=&-\bar{B}, 
	\end{align}
\end{subequations}
where $r$ is the potential and $\bar{r}$ denotes the complex conjugate of $r$.
The FONLS equation $(\ref{e1})$ emerges from the compatibility condition $U_t-V_x+[U,V]=0$ of the above pair of the linear equations $(\ref{e2})$ and $(\ref{e3})$.

In the literature, numerous efforts have been made to construct RW solutions of nonlinear partial  differential equations using DT technique \cite{darboux}.  The one-fold Darboux transformation for the Eq. (\ref{e1}) is given by \cite{feng}
\begin{equation}
\hat{r}(x,t)=r(x,t)+\frac{2(\lambda_1+\bar{\lambda}_1)f_1 \bar{g}_1}{|f_1|^2+|g_1|^2},
\label{e7}
\end{equation}
where $\varphi=(f_1(x,t), g_1(x,t))^T$ is a non-zero solution of the first order Lax pair Eqs. $(\ref{e2})$ and $(\ref{e3})$ at $\lambda=\lambda_1$, $r(x,t)$ and $\hat{r}(x,t)$ denotes the seed and first iterated solution of Eq. (\ref{e1}). 
\section{ Periodic travelling wave solutions of (1) }
\par  We consider the following form of periodic wave solution for the function $r(x,t)$, that is
\begin{equation}
r(x,t)= R(x,t) e^{ibt},
\label{e8}
\end{equation}
with the aim that $R(x,t)=R(x-ct)$ should be determined in terms of Jacobian elliptic functions. Here $b$ and $c$ are the real parameters. Substituting Eq. (\ref{e8}) into (\ref{e1}) and separating real and imaginary parts, we obtain the following two differential equations for the function $R(x,t)$, namely
\begin{subequations}
\begin{align}
&R_{xx}+2R^2 R-b R=0,
\label{e9}\\
&cR_t+R_x+\epsilon(R_{xxxxx}+10R^2R_{xxx}+40 RR_xR_{xx}+30R^4R_x+10R^3_x)=0.
\label{e10}
\end{align}
\end{subequations}
\par Upon integrating Eq. (\ref{e9}) once, we obtain 
\begin{align}
R^2_x+R^4-b R^2-d=0,
\label{e11}
\end{align}
where $d$ is the integration constant. 
\begin{figure}[!ht]
	\begin{center}
		\begin{subfigure}{0.45\textwidth}
			\includegraphics[width=\linewidth]{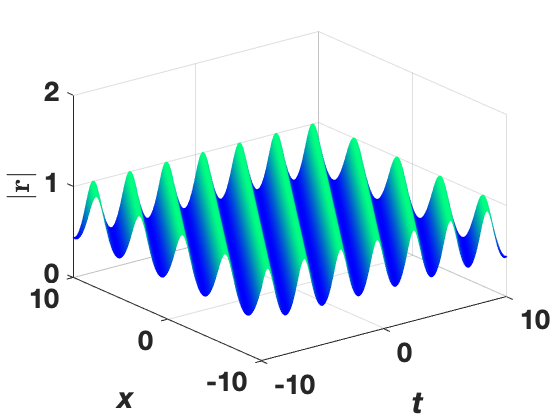}
			\caption{}
		\end{subfigure}
		\begin{subfigure}{0.45\textwidth}
			\includegraphics[width=\linewidth]{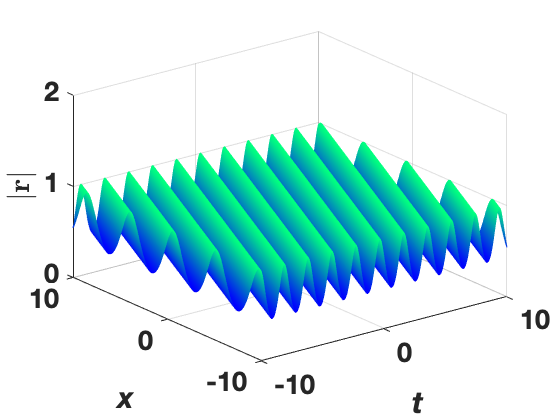}
			\caption{}
		\end{subfigure}\\
		\begin{subfigure}{0.45\textwidth}
			\includegraphics[width=\linewidth]{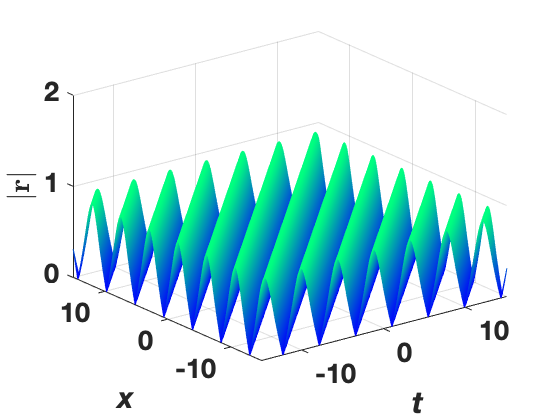}
			\caption{}
		\end{subfigure}
		\begin{subfigure}{0.45\textwidth}
			\includegraphics[width=\linewidth]{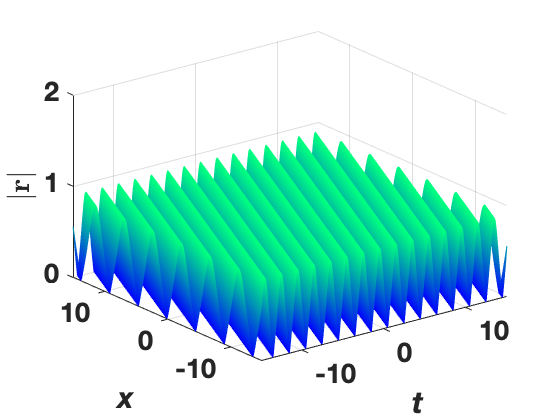}
			\caption{}
		\end{subfigure}
	\end{center}
	\vspace{-0.5cm}
	\caption{Periodic profile of (\ref{e8}) using (\ref{e12}) for $k=0.9$ with (a) $\epsilon=0.75$ and (b) $\epsilon=2.5$ and periodic profile of (\ref{e8}) using (\ref{e13}) for $k=0.9$ with (c)  $\epsilon=0.75$ and (d) $\epsilon=2.5$. }
	\label{dnnfig2}
\end{figure}

\par A compatible solution for the Eqs. (\ref{e9}) and (\ref{e10}) can be obtained with the help of (\ref{e11}) in terms of Jacobian elliptic functions with suitable choice of parameters. Among the two types of elliptic function solutions which it admits, one turns out to be the positive-definite $dn$-periodic wave 
\begin{align}
&R(x, t)=dn(x-c t,k) ,\quad b=2-k^2,\quad c=\sqrt{\epsilon(6-6k^2+k^4)+1
},\quad d=-(1-k^2),
\label{e12}
\end{align}
and the other one is sign-indefinite $cn$-periodic wave,
\begin{align}
&R(x, t)=k~ cn(x-c t,k),\quad b=2k^2-1,\;\;  c=\sqrt{\epsilon(1-6k^2+6k^4)+1
}, \;\;\;  d=k^2(1-k^2),
\label{e13}
\end{align}
where the parameter $k\in(0,1)$ is the elliptic modulus. It is staightforward to verify that (\ref{e12}) and (\ref{e13}) constitutes the solution of Eq. (\ref{e1}).

In Fig. \ref{dnnfig2} we show the qualitative profiles of $dn$ and $cn$-periodic waves of (\ref{e1}). To draw these figures, we fix the modulus parameter $k$ as $0.9$ and vary the $\epsilon$ value.  In Figs. \ref{dnnfig2}(a)-(b), we plot the solution (\ref{e8}) with (\ref{e12}) for two different values of $\epsilon$.  In Figs. \ref{dnnfig2}(c)-(d), the $cn$-periodic profiles of (\ref{e8}) with (\ref{e13}) are shown for $\epsilon=0.75$ and $\epsilon=2.5$.  From these figures, we notice that the periodicity of the $dn$ and $cn$ waves increases in the $(x-t)$ plane (see Figs. \ref{dnnfig2}(b) and \ref{dnnfig2}(d)) when we increase the value of the system parameter ($\epsilon$) from $0.75$ to $2.5$. 
\subsection{Nonlinearization of Lax pair}
In the one-fold DT formula (\ref{e7}), we intend to feed the elliptic function solutions, (\ref{e12}) and (\ref{e13}), as seed solutions. To do this we should know for what eigenvalues $(\lambda_1)$ these solutions arise. Conventionally, in the DT method, the eigenvalues can be determined by substituting the considered seed solution into the Lax pair equations and solving the underlying first order ODEs (in our case four equations). However, in the present case, it is very difficult to integrate the Lax pair equations with the presence of Jacobian elliptic functions. To overcome this difficulty we seek another technique, namely nonlinearization of Lax pair \cite{zhou, zhou1} and identify the eigenvalues $(\lambda_1)$ that correspond to the solutions (\ref{e12}) and (\ref{e13}).

\par  In this method, by introducing a Bargmann constraint and appropriately using the Lax pairs given in (\ref{e2}) and (\ref{e3}), we derive two differential constraints. Comparing these two differential constraints with Eqs. (\ref{e9}) and (\ref{e11}) we determine the admissible eigenvalues and the corresponding squared eigenfunctions.

  We consider the constraint between the solution of Eq. (\ref{e1}) and the squared eigenfunctions in the form
\begin{align}
r(x,t)=f^2_1+\bar{g}^2_1,
\label{e14}
\end{align}
where $(f_1,g_1)^T$ is a non-zero solution of the linear system of Eqs. (\ref{e2}) and (\ref{e3}) at $\lambda=\lambda_1$. Substituting Eq. (\ref{e14}) into Eq. (\ref{e2}), we identify that the underlying equations can be written as a finite-dimensional Hamiltonian system, that is
\begin{align}
\frac{df_1}{dx}=\frac{\partial H_0}{\partial g_1},\qquad \frac{dg_1}{dx}=-\frac{\partial H_0}{\partial f_1},
\label{e15}
\end{align}
with $H_0$ is given by
\begin{align}
H_0=\lambda_1 f_1g_1+\bar{\lambda}_1\bar{f}_1 \bar{g}_1+\frac{1}{2}(f^2_1+\bar{g}^2_1)(\bar{f}^2_1+g^2_1).
\label{e16}
\end{align}
Differentiating Eq. (\ref{e14}) with respect to $x$ and using the Eqs. (\ref{e2}) and (\ref{e16}), we can obtain the following first order ODE, that is
\begin{align}
r_x+2irF_0=2(\lambda_1 f^2_1-\bar{\lambda}_1 \bar{g}^2_1),
\label{e20}
\end{align}
where $F_0=i(f_1g_1-\bar{f}_1\bar{g}_1)$.

Differentiating Eq. (\ref{e20}) with respect to $x$, and using (\ref{e16}) in it, we end up at
\begin{align}
r_{xx}+2i(F_0+i\lambda_1-i\bar{\lambda}_1)r_x+2|r|^2r=4(|\lambda_1|^2+H_0+i F_0(\lambda_1-\bar{\lambda}_1))r,
\label{e211}
\end{align} 
where $F_1=\lambda_1 f_1g_1+\bar{\lambda}_1\bar{f}_1 \bar{g}_1+\frac{1}{2}(|f_1|^2+|g_1|^2)^2$. One can unambiguously verify that $\frac{dF_0}{dx}=0$ and $\frac{dF_1}{dx}=0$.  These two constants of motion ($F_0$ and $F_1$) can be connected to $H_0$ (given in (\ref{e16})) through the relation  $H_0=F_1-\frac{1}{2}F^2_0$.  

The second-order Eq. (\ref{e211}) admits the following Lax representation \cite{chen2}, that is 
\begin{align}
\frac{d}{dx}J(\lambda)=[U(\lambda,r),J(\lambda)],\qquad \lambda\in \mathbb{C},
\label{ee1}
\end{align}
where $U(\lambda,r )$ and $r(x,t)$ are given in (\ref{e2}) and (\ref{e14}) respectively, and the matrix $J(\lambda)$ is defined by, 
\begin{align}
J(\lambda)=\begin{bmatrix}
J_{11}(\lambda) & J_{12}(\lambda)  \\
\overline{J}_{12}(-\lambda)& -\overline{J}_{11}(-\lambda)
\end{bmatrix}.
\label{ee2}
\end{align}
The components $J_{11}$ and $J_{12}$ read $J_{11}(\lambda)=1- \left(\frac{f_{1} g_{1}}{\lambda-{\lambda}_1}-\frac{\bar{f}_{1} \bar{g}_{1}}{\lambda+{\bar{\lambda}}_1}\right)$ and $J_{12}(\lambda)=
\left(\frac{f^2_{1} }{\lambda-{\lambda}_1}+\frac{ \bar{g}^2_{1}}{\lambda+{\bar{\lambda}}_1}\right)$. With the help of Eqs. (\ref{e14}) and (\ref{e20}),  the functions  $J_{11}(\lambda)$ and $J_{12}(\lambda)$ can be rewritten as
\begin{align}
J_{11}(\lambda)&=\frac{(\lambda-\lambda_1)(\lambda+\bar{\lambda}_1)+i F_0(\lambda-\lambda_1+\bar{\lambda}_1)+\frac{1}{2}(F^2_0+|r|^2)-F_1}{(\lambda-\lambda_1)(\lambda+\bar{\lambda}_1)},\nonumber\\
J_{12}(\lambda)&=
\frac{(\lambda-\lambda_1+\bar{\lambda}_1+iF_0)r+\frac{r_x}{2}}{(\lambda-\lambda_1)(\lambda+\bar{\lambda}_1)}.
\label{ee3}
\end{align}

One can verify that the upper right component $(J_{12})$ that appear in the matrix Eq. $(\ref{ee1})$,  yields the same second-order ODE (\ref{e211}).

 We consider the eigenvalue in the form $\lambda_1=\xi+i \eta$, where $\xi$ and $\eta$ are two real parameters. With this choice Eq. (\ref{e211}) can be rewritten in the form
\begin{align}
r_{xx}+2i(F_0-2\eta)r_x+2|r|^2r=4(\xi^2+\eta^2+F_1-\frac{1}{2}F^2_0+2\eta F_0)r.
\label{e21}
\end{align} 

While evaluating the determinant of $J(\lambda)$, we notice that the determinant contains two simple poles which can be expressed in terms of $F_0$ and $F_1$. Imposing the constraint $\overline{J}_{11}(-\lambda)=J_{11}(\lambda)$ and using the exact forms given in Eq. (\ref{ee3}) for $J_{11}(\lambda)$ and $J_{12}(\lambda)$ we obtain the following expression for $det J(\lambda)$: 
\begin{align}
\text{det}J(\lambda)=&-\Big[\frac{(\lambda-\lambda_1)(\lambda+\bar{\lambda}_1)+i F_0(\lambda-\lambda_1+\bar{\lambda}_1)+\frac{1}{2}(F^2_0+|r|^2)-F_1}{(\lambda-\lambda_1)(\lambda+\bar{\lambda}_1)}\Big]^2\nonumber\\
&+\frac{[(\lambda-\lambda_1+\bar{\lambda}_1+iF_0)r+\frac{1}{2}r_x][(\lambda-\lambda_1+\bar{\lambda}_1+iF_0)\bar{r}-\frac{1}{2}\bar{r}_x]}{(\lambda-\lambda_1)^2(\lambda+\bar{\lambda}_1)^2}.
\label{Ee1}
\end{align}

The det $J(\lambda)$ has double poles, one at $\lambda=\lambda_1$ and another at $\lambda=-\bar{\lambda}_1$. By removing the poles and making appropriate simplifications, we can identify the following two differential constraints, namely 
\begin{align}
\label{e22}
\bar{r}r_x-r\bar{r}_x=& 2i(2\eta-F_0)|r|^2+2iF_0(F^2_0+2\eta F_0-2F_1), \\
|r_x|^2+|r|^4=&4(\xi^2+\eta^2-\frac{1}{2}F^2_0+F_1-2\eta F_0)|r|^2+4\xi^2F^2_0\nonumber\\
&-(F^2_0+2\eta F_0-2F_1)(5F^2_0-2\eta F_0-2F_1).
\label{e23}
\end{align}
Substituting the expression $r(x,t)$ (given in Eq. (\ref{e8})) in (\ref{e22}) the left-hand side  becomes zero for both the periodic waves. The right-hand side of Eq. (\ref{e22}) yields
\begin{align}
F_0=2\eta,\quad \eta(F_1-4\eta^2)=0.
\label{e24}
\end{align}
Upon comparing the Eqs. (\ref{e21}) and (\ref{e23}) with the expressions (\ref{e9})  and (\ref{e11}) respectively, we can fix the parameters $b$ and $d$ in the form
\begin{align}
 b=4(\xi^2-5\eta^2+F_1),\quad d=4[4\xi^2\eta^2-(F_1-4\eta^2)(F_1-8\eta^2)].
\label{e25}
\end{align}

Equation (\ref{e24}) yields two conditions, namely (i) $\eta=0$ and (ii) $F_1=4\eta^2$  $(\eta\neq 0)$.  In the first choice, we find $F_0=0$ and hence the parameters $b$ and $d$ in Eq. (\ref{e25}) assume the form
\begin{align}
 b=4(\xi^2+F_1), \quad d=-4F^2_1.
\label{e26}
\end{align}
In the second choice $(\eta\neq 0)$, the parameters $b$ and $d$ take the value
\begin{align}
 b=4(\xi^2-\eta^2), \quad d=16\xi^2\eta^2.
\label{e27}
\end{align}

While the parameter $d$ has a sign difference in these two cases the  parameter $b$ takes the same sign.  Upon comparing the values of the parameters which we found through nonlinearization of Lax pair technique (Eqs. (\ref{e26}) and (\ref{e27})) with the ones obtained through travelling wave reduction (Eqs. (\ref{e12}) and (\ref{e13})), we conclude that the minus sign case represents the $dn$ periodic wave whereas the plus sign case corresponds to the $cn$ periodic wave. In the following, we determine the admissible eigenvalues and eigenfunctions of the periodic wave solutions (\ref{e12}) and (\ref{e13}) of Eq. (\ref{e1}).
\subsection{Eigenvalues and periodic Eigenfunctions} 
Upon comparing the Eqs. (\ref{e12}) and (\ref{e26}) with $\lambda_1=\xi$ and $F_1=\pm \frac{1}{2}\sqrt{1-k^2}$, we obtain two real eigenvalues for the $dn$-periodic wave, namely
\begin{align}
\lambda_{\pm}=\frac{1}{2}(1\pm \sqrt{1-k^2}).
\label{e28}
\end{align}
As far as the $cn$-periodic wave is concerned we compare the Eqs. (\ref{e13}) and (\ref{e27}) with $\lambda_1=\xi+i\eta$.  Here, we find
\begin{align}
\lambda_{\pm}=\frac{1}{2}(k\pm i\sqrt{1-k^2}).
\label{e29}
\end{align}

Upon solving the Eqs. (\ref{e14}) and (\ref{e20}) for the squared eigenfunctions $f^2_1$ and $\bar{g}^2_1$, we obtain the following expressions, namely
\begin{align}
f^2_1=\frac{2\lambda_1 r+r_x}{2(\lambda_1+\bar{\lambda}_1)},\quad \bar{g}^2_1=\frac{2\bar{\lambda}_1 r-r_x}{2(\lambda_1+\bar{\lambda}_1)}.
\label{e30}
\end{align}

Next, we determine the explicit form of the product of the eigenfunctions $f_1$ and $g_1$ in terms of $r(x,t)$.  For the $dn$-periodic case, we already know $\eta=0$ and $F_0=0$. From the identity $H_0=F_1-\frac{1}{2}F^2_0$ we can fix $H_0=F_1=\pm \frac{1}{2}\sqrt{1-k^2}$.  Substituting these relations back in Eq. (\ref{e16}) along with $|r|^2=|f_1^2+\bar{g}^2_1|$ and $\xi=\lambda_1=\lambda_+$, we obtain
  \begin{align}
 f_1g_1=-\frac{1}{4\lambda_1}[|r(x,t)|^2+\sqrt{1-k^2}],
 \label{e31}
 \end{align}
where $r(x,t)=R(x,t)\e^{ibt}$ in which $R(x,t)$ is given in Eq. (\ref{e12}).

As far as the $cn$-periodic wave case is concerned, we identify $H=2\eta^2$ from the relations  $F_0=2\eta$ and $F_1=4\eta^2$.  Substituting these functions back in Eq. (\ref{e16}) with $\lambda_1=\xi+i\eta$,  we find
\begin{align}
f_1g_1=-\frac{1}{2k}[|r(x,t)|^2+ik\sqrt{1-k^2}],
\label{e32}
\end{align} 
where $r(x,t)=R(x,t)\e^{ibt}$ and $R(x,t)$ is given in Eq. (\ref{e13}).   Upon substituting the obtained eigenvalue $(\lambda_1)$, periodic eigenfunctions $(f_1^2,\bar{g}^2_1,f_1g_1)$ and periodic wave solutions $r(x,t)$ in the one-fold DT formula $(\ref{e7})$, we can create the periodic background. As our aim is to construct RW on the top of this periodic waves, we move on to construct a second linearly independent solution of the spectral problem (\ref{refno}) for the same eigenvalue $\lambda=\lambda_1$ which in turn provides the desired result.
\section{RW solutions on the periodic background }
We construct a second linearly independent solution ($\varphi=(\hat{f}_1,\hat{g}_1)^T$) to the Eq. (\ref{e1}) with the following two properties:  (i)  the second solution $\varphi=(\hat{f}_1,\hat{g}_1)^T$ should also posseses the same eigenvalue $\lambda=\lambda_1$ and (ii) it should exhibit a non-periodic localized profile.  Based on these two requirements, we choose the second linearly independent solution to the spectral problem (\ref{e2}) in the form 
\begin{align}
\label{e33}
\hat{f}_1 = f_1 \delta_1-\frac{2\bar{g}_1}{|f_1|^2+|g_1|^2}, \quad \hat{g}_1 =  g_1 \delta_1+\frac{2\bar{f}_1}{|f_1|^2+|g_1|^2},
\end{align}
where $\delta_1(x,t)$ is an unknown function which is to be determined.   
By inserting Eq. (\ref{e33}) into Eqs. (\ref{e2}) and (\ref{e3}), we obtain the following two first-order partial differential equations for the unknown function $\delta_1$, that is 
\begin{align}
\label{e35}
\frac{\partial \delta_1}{\partial x} = M_1:= & -\frac{4(\lambda_1+\bar{\lambda}_1)\bar{f}_1\bar{g}_1}{\left(|\hat{f}_1|^2+|\hat{g}_1|^2\right)^2}, \\
\label{e36}
\frac{\partial \delta_1}{\partial t} = M_2: = & \frac{4(\bar{f}^2_1S_1+2\bar{g}^2_1S_2-\bar{f}_1\bar{g}_1S_3)}{(|f_1|^2+|g_1|^2)^2},
\end{align}
where 
\begin{align}
S_1=~ & (\lambda_1-\bar{\lambda}_1)(4\epsilon(\lambda_1+\bar{\lambda}_1)|r|^2r+r(i+6\epsilon r_x\bar{r}+8\epsilon(\lambda_1+\bar{\lambda}_1)(\lambda^2_1+\bar{\lambda}^2_1))\nonumber\\&+\epsilon(r_{xxx}+2r_{xx}(\lambda_1+\bar{\lambda}_1)+4r_x(\lambda^2_1-|\lambda_1|^2+\bar{\lambda}^2_1))),\nonumber\\
S_2=~& (\lambda_1-\bar{\lambda}_1)(4\epsilon(\lambda_1+\bar{\lambda}_1)|r|^2\bar{r}+\bar{r}(-i+6\epsilon \bar{r}_xr+8\epsilon(\lambda_1+\bar{\lambda}_1)(\lambda^2_1+\bar{\lambda}^2_1))\nonumber\\&+\epsilon(\bar{r}_{xxx}+2\bar{r}_{xx}(\lambda_1+\bar{\lambda}_1)+4\bar{r}_x(\lambda^2_1-|\lambda_1|^2+\bar{\lambda}^2_1))),\nonumber\\
S_3=~&(\lambda_1+\bar{\lambda}_1)(1+2i(\lambda_1-\bar{\lambda}_1)+2\epsilon(3|r|^4+\bar{r}r_{xx}-r_x(\bar{r}_x-2\bar{r}(\lambda_1-\bar{\lambda}_1))\nonumber\\&+8(\lambda^4_1-|\lambda_1|^2(\lambda^2_1+\bar{\lambda}^2_1)+|\lambda_1|^4+\bar{\lambda}^4_1)+r(\bar{r}_{xx}-2\bar{r}_x(\lambda_1-\bar{\lambda}_1)+4\bar{r}\nonumber\\&\times(\lambda^2_1-|\lambda_1|^2+\bar{\lambda}^2_1))))\nonumber.
\end{align}

\begin{figure}[!ht]
	\begin{center}
		\begin{subfigure}{0.43\textwidth}
			\includegraphics[width=\linewidth]{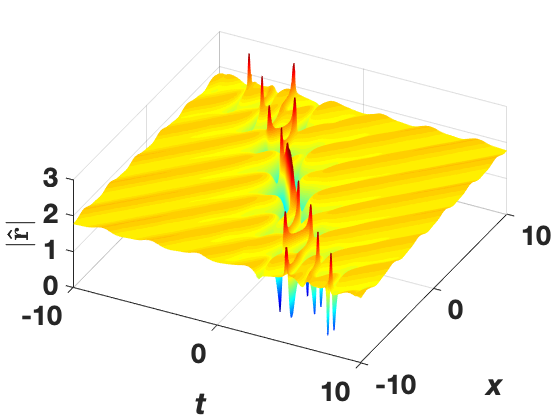}
			\caption{}
		\end{subfigure}
		\begin{subfigure}{0.43\textwidth}
			\includegraphics[width=\linewidth]{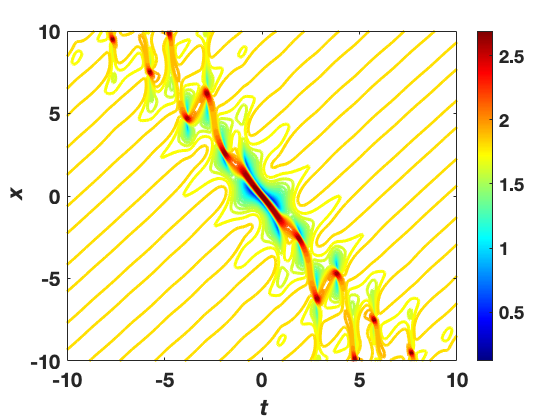}
			\caption{}
		\end{subfigure}\\
		\begin{subfigure}{0.43\textwidth}
			\includegraphics[width=\linewidth]{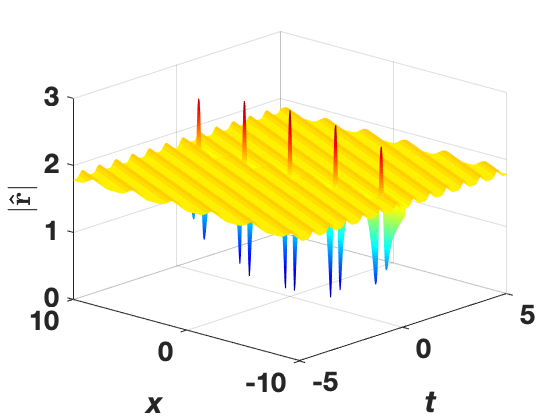}
			\caption{}
		\end{subfigure}
		\begin{subfigure}{0.43\textwidth}
			\includegraphics[width=\linewidth]{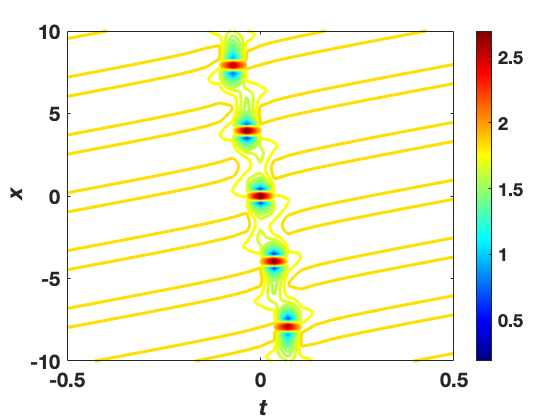}
			\caption{}
		\end{subfigure}\\
		\begin{subfigure}{0.43\textwidth}
			\includegraphics[width=\linewidth]{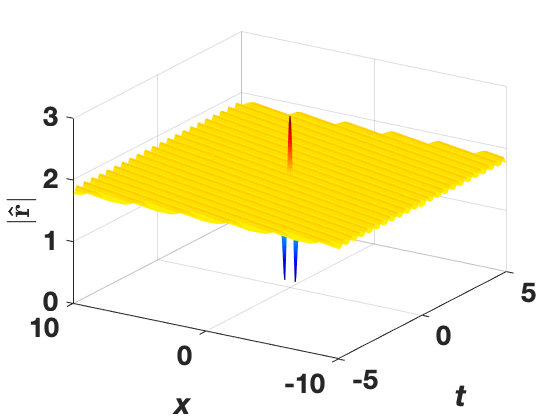}
			\caption{}
		\end{subfigure}
		\begin{subfigure}{0.43\textwidth}
			\includegraphics[width=\linewidth]{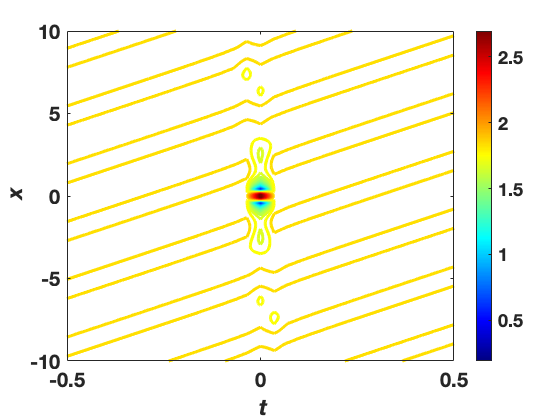}
			\caption{}
		\end{subfigure}
	\end{center}
	\vspace{-0.7cm}
	\caption{Rogue dn-periodic wave profile of (\ref{e39}) with (\ref{e12}) and $k=0.6$ for three different values (a)-(b) $\epsilon=0.02$, (c)-(d) $\epsilon=4.5$ and (e)-(f) $\epsilon=15$. Panels (b), (d) and (f) are the corresponding contour plots of (a), (c) and (e)}.
	\label{dnnfig3}
\end{figure}
\begin{figure}[!ht]
	\begin{center}
		\begin{subfigure}{0.45\textwidth}
			\includegraphics[width=\linewidth]{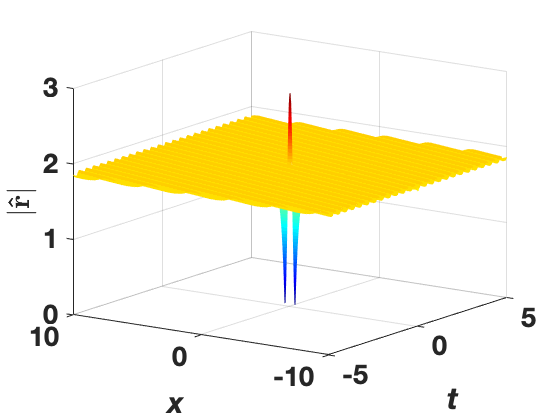}
			\caption{}
		\end{subfigure}
		\begin{subfigure}{0.45\textwidth}
			\includegraphics[width=\linewidth]{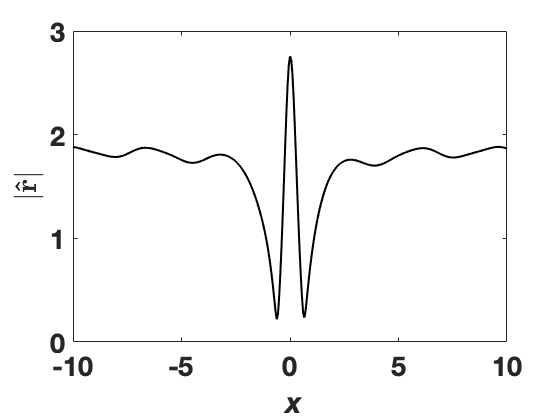}
			\caption{}
		\end{subfigure}\\
		\begin{subfigure}{0.45\textwidth}
			\includegraphics[width=\linewidth]{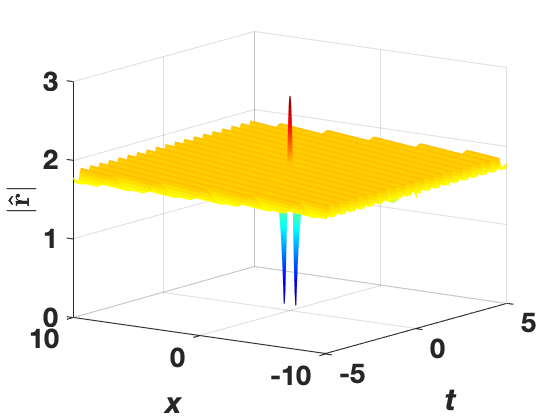}
			\caption{}
		\end{subfigure}
		\begin{subfigure}{0.45\textwidth}
			\includegraphics[width=\linewidth]{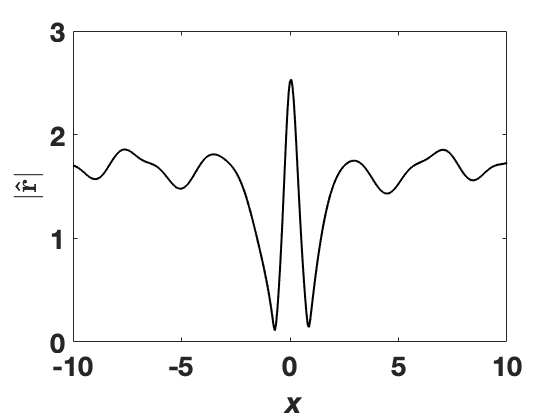}
			\caption{}
		\end{subfigure}\\
		\begin{subfigure}{0.45\textwidth}
			\includegraphics[width=\linewidth]{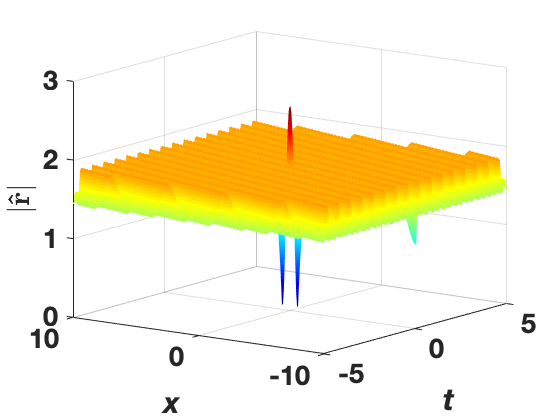}
			\caption{}
		\end{subfigure}
		\begin{subfigure}{0.45\textwidth}
			\includegraphics[width=\linewidth]{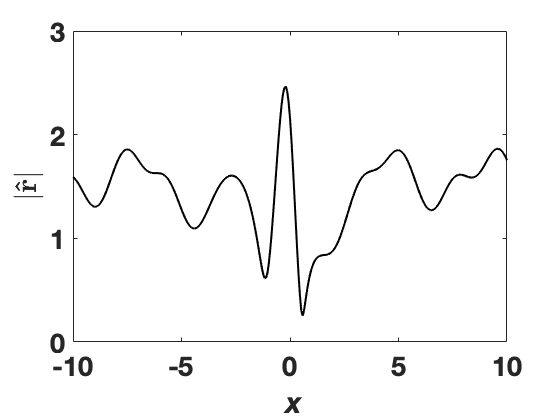}
			\caption{}
		\end{subfigure}
	\end{center}
	\vspace{-0.7cm}
	\caption{Rogue dn-periodic wave profile of (\ref{e39}) with (\ref{e12}) and $\epsilon=29$, (a)-(b) $k=0.5$, (c)-(d) $k=0.7$  and (e)-(f) $k=0.9$. Panels (b), (d) and (f) are the  corresponding two dimensional plots of (a), (c) and (e). }
	\label{dnnfig31}
\end{figure}
\begin{figure}[!ht]
	\begin{center}
		\begin{subfigure}{0.45\textwidth}
			\includegraphics[width=\linewidth]{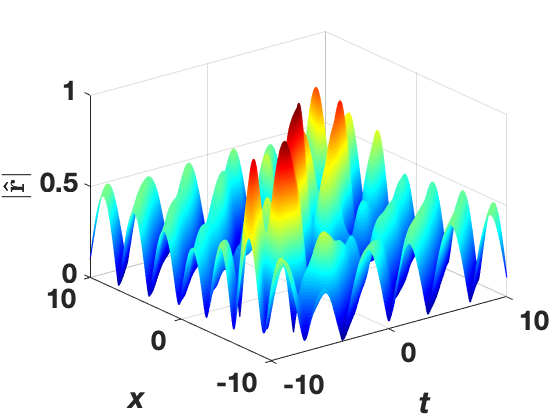}
			\caption{}
		\end{subfigure}
		\begin{subfigure}{0.45\textwidth}
			\includegraphics[width=\linewidth]{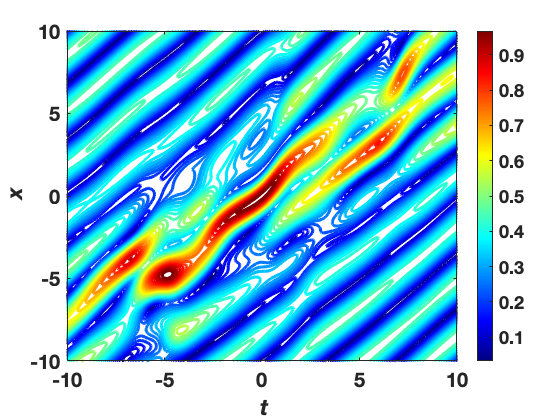}
			\caption{}
		\end{subfigure}\\
		\begin{subfigure}{0.45\textwidth}
			\includegraphics[width=\linewidth]{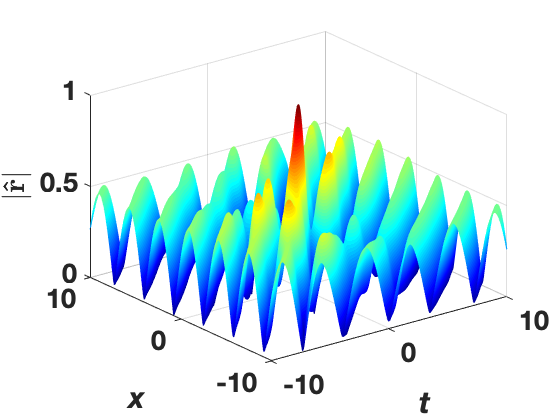}
			\caption{}
		\end{subfigure}
		\begin{subfigure}{0.45\textwidth}
			\includegraphics[width=\linewidth]{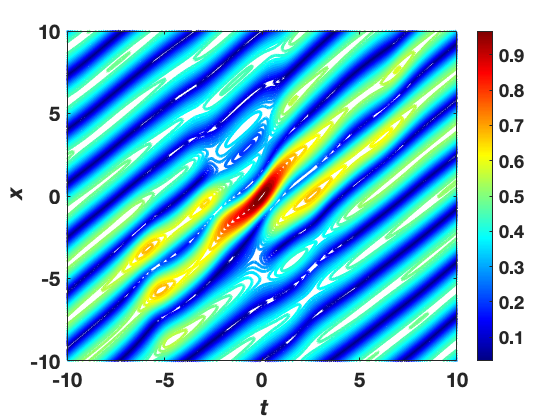}
			\caption{}
		\end{subfigure}\\
		\begin{subfigure}{0.45\textwidth}
			\includegraphics[width=\linewidth]{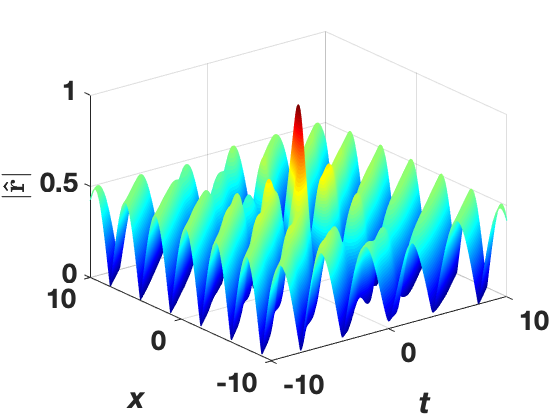}
			\caption{}
		\end{subfigure}
		\begin{subfigure}{0.45\textwidth}
			\includegraphics[width=\linewidth]{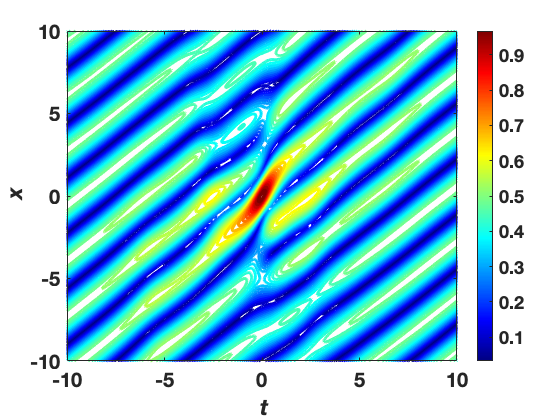}
			\caption{}
		\end{subfigure}
	\end{center}
	\vspace{-0.7cm}
	\caption{{ Rogue $cn$-periodic wave profile of (\ref{e39}) with (\ref{e13}) and $k=0.5$ for three different values (a)-(b) $\epsilon=0.2$, (c)-(d) $\epsilon=0.65$ and (e)-(f) $\epsilon=1.2$. Panels (b), (d) and (f) are corresponding contour plots of (a), (c) and (e).}}
	\label{dnnfig4}
\end{figure}

The system of partial differential equations (\ref{e35}) and (\ref{e36}) are compatible with each other ($M_{1t}=M_{2x}$) because both the expressions are derived from the compatible Lax Eqs. (\ref{e2}) and (\ref{e3}). These two expressions (\ref{e35}) and (\ref{e36}) can be solved  with the integration formula   
\begin{align}
\label{e37}
\delta_1(x,t)= & \int_{x_0}^{x} M_1(x',t)dx'+ \int_{t_0}^t M_2(x_0,t')dt',
\end{align}
where $(x_0,t_0)$ is arbitrarily fixed. The presence of higher-order derivative terms in (\ref{e37}), enforces us to integrate the above underlying integrals numerically using the Newton-Raphson method.
 \par Substituting the considered second seed solution $\varphi =(\hat{f}_1,\hat{g}_1)^T$ of the linear equations (\ref{e2})-(\ref{e3}) with $\lambda=\lambda_1$ in the one-fold DT formula (\ref{e7}), we obtain a new solution to the FONLS Eq. (\ref{e1}) of the form
\begin{align}
\label{e39}
\hat{r}(x,t)= & r(x,t)+\frac{2(\lambda_1+\bar{\lambda}_1)\hat{f_1}\bar{\hat{g}}_1}{|\hat{f}_1|^2+|\hat{g}_1|^2}\nonumber\\
	=&r(x,t)+\frac{2(\lambda_1+\bar{\lambda}_1)[f_1(|f_1|^2+|g_1|^2)\delta_1-2\bar{g}_1][\bar{g}_1(|f_1|^2+|g_1|^2)\bar{\delta}_1+2f_1]}{|f_1(|f_1|^2+|g_1|^2)\delta_1-2\bar{g}_1|^2+|\bar{g}_1(|f_1|^2+|g_1|^2)\bar{\delta}_1+2f_1|^2},
\end{align}
where $\hat{f}_1$ and $\hat{g}_1$ are given in (\ref{e33}), the periodic wave solution $r(x,t)$ can be taken from (\ref{e8}) in which $R(x,t)$ is given in (\ref{e12}) and (\ref{e13}).  If we consider the seed solution $r(x,t)$ in $dn$ periodic wave form with $\lambda_1=\frac{1}{2}[1+\sqrt{1-k^2}]$, then the new solution reveals RW structure on the $dn$ periodic wave background. Similarly, if we consider the seed solution in $cn$ periodic wave form with $\lambda_1=\frac{1}{2}[k+i\sqrt{1-k^2}]$, then the new solution creates a RW structure on the $cn$ periodic wave background.

The surface plots of the rogue periodic wave solution on the $dn$ periodic wave background are shown in Figs. \ref{dnnfig3}(a)-\ref{dnnfig3}(e) using (\ref{e39}) with $\lambda=\frac{1}{2}(1+ \sqrt{1-k^2})$ and $k=0.6$.  Typical distributions of $|\hat{r}|$ show how the nature of RW structures get modified in the elliptic function background when we vary the system parameter $\epsilon$.  The RW attains its highest amplitude at its origin, that is $(x_0, t_0)=(0, 0)$. In Fig. \ref{dnnfig3}(a), we observe that the maximum amplitude of periodic RW is $|\hat{r}|=2.781$ for $\epsilon=0.02$ . The corresponding contour plot is presented in Fig. \ref{dnnfig3}(b). Figures \ref{dnnfig3}(c) and \ref{dnnfig3}(e) represent the qualitative nature of RWs on $dn$-periodic background for two other values of the system parameter, namely $\epsilon=4.5$ and $15$, respectively.  The corresponding contour plots are displayed in Figs. \ref{dnnfig3}(d) and \ref{dnnfig3}(f). We notice that the RW retains its height ($|\hat{r}|=2.781$) while we increase the value of the parameter $\epsilon$ from $0.02$ to $15$. A main difference which we observe here is that when we vary the system parameter $\epsilon$ the orientation of the localized waves changes and the frequency of the periodic background waves increases in the $(x-t)$ plane.

We move on to investigate the RWs on $dn$-periodic background when we alter the elliptic function modulus value $k$.  In this investigation we fix all the parameter values be the same as in the above investigation and we only vary the elliptic modulus parameter value of $k$.  For the parametric value ($\epsilon=29$), the surface plots of $|\hat{r}|$ of RWs on the $dn$ periodic wave background with three different values of elliptic modulus ($k=0.5,~0.7$ and $0.9$) are presented in Figs. \ref{dnnfig31}(a), \ref{dnnfig31}(c) and \ref{dnnfig31}(e), respectively. The corresponding two dimensional plots are presented in Figs. \ref{dnnfig3}(b), \ref{dnnfig3}(d) and \ref{dnnfig3}(f).  In these figures, we find that the amplitude of RWs reaches a maximum value at their origin. The amplitude of the RWs differs in each case and it is found that $|\hat{r}|=2.823,~2.729$ and $2.591$, respectively for $k=0.5,~k=0.7$ and $0.9$. Here we notice that the amplitude of the RW decreases when we enhance the $k$ value from $k=0.5$ to $0.9$.  However, the frequency of the periodic background waves increases when we increase the values of $\epsilon$ and $k$.  In the 2D plots, one can clearly visualize  the changes that occur in frequency and amplitude of the $dn$- periodic wave around the RW. From the outcome, we conclude that when we increase the $k$ value the amplitude of the RW decreases.

\begin{figure}[!ht]
	\begin{center}
		\begin{subfigure}{0.45\textwidth}
			\includegraphics[width=\linewidth]{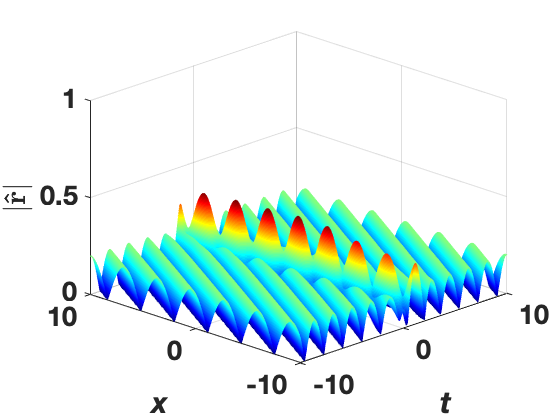}
			\caption{}
		\end{subfigure}
		\begin{subfigure}{0.45\textwidth}
			\includegraphics[width=\linewidth]{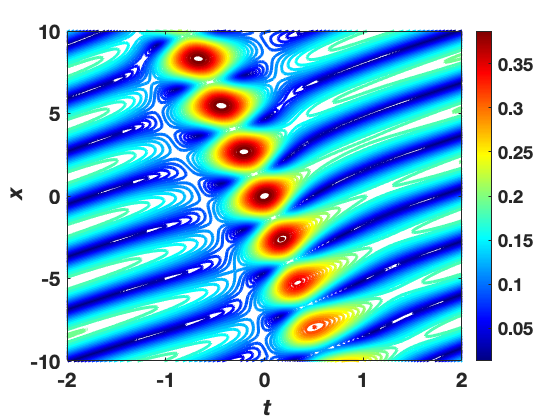}
			\caption{}
		\end{subfigure}\\
		\begin{subfigure}{0.45\textwidth}
			\includegraphics[width=\linewidth]{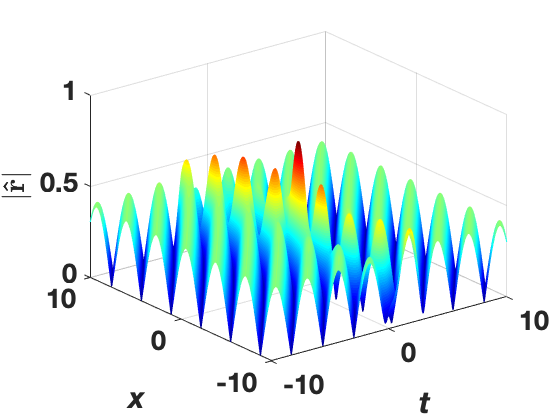}
			\caption{}
		\end{subfigure}
		\begin{subfigure}{0.45\textwidth}
			\includegraphics[width=\linewidth]{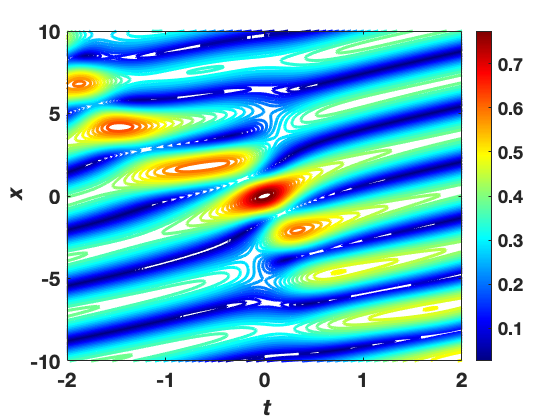}
			\caption{}
		\end{subfigure}\\
		\begin{subfigure}{0.45\textwidth}
			\includegraphics[width=\linewidth]{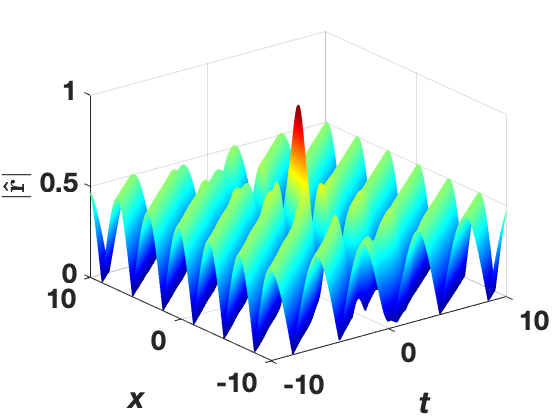}
			\caption{}
		\end{subfigure}
		\begin{subfigure}{0.45\textwidth}
			\includegraphics[width=\linewidth]{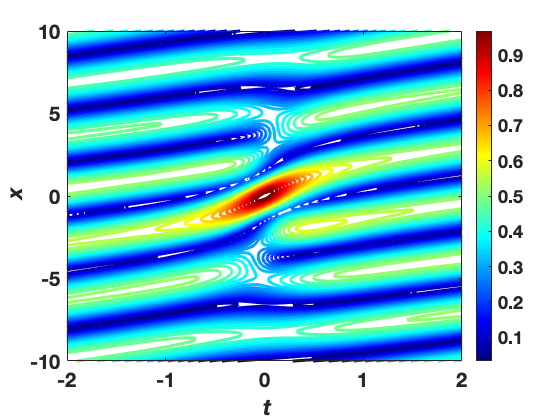}
			\caption{}
		\end{subfigure}
	\end{center}
	\vspace{-0.6cm}
	\caption{{ Rogue $cn$-periodic wave profile of (\ref{e39}) with (\ref{e13}) and $\epsilon=2.5$ for three different values (a)-(b) $k=0.2$, (c)-(d) $k=0.4$ and (e)-(f) $k=0.5$. Panels (b), (d) and (f) are corresponding contour plots of (a), (c) and (e)}}
	\label{dnnfig41}
\end{figure}

Now, we investigate how the RW structures evolve in the $cn$-periodic background for the FONLS equation.  Figure \ref{dnnfig4} shows the surface plots of the RWs on the $cn$ periodic wave background  using the solution (\ref{e39}) with $\epsilon=0.2$, $\epsilon=0.65$, $\epsilon=1.2$ and $k=0.5$ and $\lambda=\frac{1}{2}(k+ i\sqrt{1-k^2})$.  The maximum amplitude of RWs is found to be $|\hat{r}|=1.0$.  This result is demonstrated in Figs. \ref{dnnfig4}(a), \ref{dnnfig4}(c) and \ref{dnnfig4}(e).  The corresponding contour plots are given in the right column in Figs. \ref{dnnfig4}(b), \ref{dnnfig4}(d) and \ref{dnnfig4}(f). The amplitude of RWs retains its height in all three cases. 

\begin{figure}[ht]
		\begin{subfigure}[c][1\width]{0.32\textwidth}
			\centering
			\includegraphics[width=\linewidth]{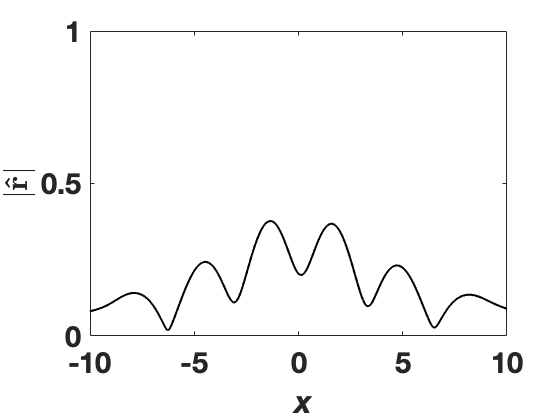}
			\caption{}
		\end{subfigure}
		\hfill 	
		\begin{subfigure}[c][1\width]{0.32\textwidth}
			\centering
			\includegraphics[width=\linewidth]{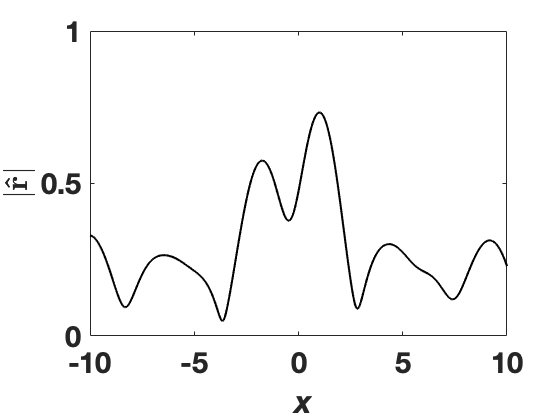}
			\caption{}
		\end{subfigure}
		\hfill	
		\begin{subfigure}[c][1\width]{0.32\textwidth}
			\centering
			\includegraphics[width=\linewidth]{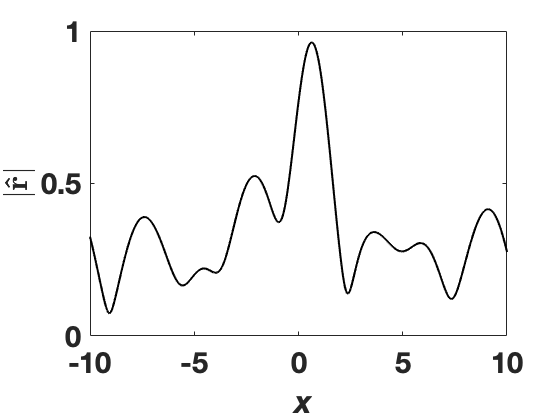}
			\caption{}
		\end{subfigure}
		\vspace{-0.2cm}
	\caption{{ Two dimensional plots for rogue $cn$-periodic wave profile of (\ref{e39}) with (\ref{e13}) and $\epsilon=2.5$ for three different values (a) $k=0.2$, (b) $k=0.4$ and (c) $k=0.5$. }}
	\label{dnnfig1}
\end{figure}

A similar dynamical characteristics is also observed while we vary the $k$ value ($k=0.2$, $k=0.4$ and $0.5$) which is demonstrated in Figs. \ref{dnnfig41}(a), \ref{dnnfig41}(c) and \ref{dnnfig41}(e) and corresponding contour plots are represented in Figs. \ref{dnnfig41}(b), \ref{dnnfig41}(d) and \ref{dnnfig41}(f), respectively.  We observe the following features when we change the elliptic modulus value $k$.  The amplitude of RW increases $|\hat{r}|=(0.4,~0.8,~1.0)$ when we increase the $k$ value ($k=0.2,~0.4,~0.5$).  To have a better understanding on the solutions we also present 2D plots in Figs. \ref{dnnfig1}(a)-\ref{dnnfig1}(c) for the same $k$ values which we considered in 3D plots. In the 2D plots, one can clearly visualize  the changes that occur in frequency and amplitude of the $cn$- periodic wave around the RW. The amplitude of the RWs increases as seen in Figs. \ref{dnnfig41} and  \ref{dnnfig1}. From the outcome, we conclude that when we increase the $k$ value the amplitude of the RW increases.
\section{Modulation instability of the periodic waves}
The interaction between dispersive and nonlinear effects causes modulational instability (MI) in a plane wave in the presence of noise or a weak frequency-shifted signal wave \cite{ankiew,obsorne,zakharov,chen4,agrawal,hase,bdec,gxu,dep}. In the literature, the stability spectrum of elliptic solutions of the focusing NLS equation has been analyzed by several authors (see for example, Refs. \cite{bdec,gxu} and references therein).  Very recently, the linear stability analysis of RWs on doubly periodic waves was investigated in detail \cite{dep}.  It has been shown that both the $dn$ and $cn$ periodic standing waves were modulationally unstable in the presence of long wave perturbations and MI growth rates are related with the periodic wave parameters. In this section, we investigate the instability of periodic waves and compute the growth rate of both of them. 

To begin, we solve the linear equations (\ref{refno}) by considering periodic waves of the form
\begin{align}
r(x,t)=R(x-ct)e^{2ibt},
\label{i1}
\end{align}
where $b$ is a real constant.  We consider a solution to the Lax pair Eq. (\ref{refno}) in the form 
\begin{align}
\varphi_1(x,t)=\xi_1(x-c t)e^{ibt+t\Omega},\quad \varphi_2(x,t)=\xi_2(x-c t)e^{-ibt+t\Omega},
\label{i2}
\end{align}	
where $\xi_1$ and $\xi_2$ are functions of their arguments and $\Omega\in~\mathbb{C}$ is another spectral parameter. Substituting Eqs. (\ref{i1}) and (\ref{i2}) into the Lax pair equations (\ref{refno}) and rearranging the expressions, we obtain the following two equations, that is
\begin{subequations}
	\label{refno1}
	\begin{align}
	\label{i3}
	\xi_{x}=\begin{pmatrix} \lambda & R\\ -\bar{R} & -\lambda \end{pmatrix}\xi, \quad
	\Omega\xi-c\begin{pmatrix} \lambda & R\\ -\bar{R} & -\lambda \end{pmatrix}\xi=\begin{pmatrix} \tilde{A}-ib & \tilde{B}\\ -\bar{\tilde{B}} & -\tilde{A}+ib \end{pmatrix}\xi, 
	\end{align}
where $\xi=(\xi_1,\xi_2)^T$, $\bar{\tilde{B}}$ is the complex conjugate of $\tilde{B}$ and the explicit form of $\tilde{A}$ and $\tilde{B}$ are given by
\begin{align}	\tilde{A}&=16\lambda^5\epsilon+8\lambda^3\epsilon|R|^2-4\lambda^2\epsilon(R\bar{R}_x-R_x \bar{R})+2i\lambda^2+2\lambda\epsilon(R\bar{R}_{xx}+\bar{R}R_{xx}-|R_x|^2\nonumber\\&+3|R|^4)
+\lambda+\epsilon(\bar{R}R_{xxx}-R\bar{R}_{xxx}+R_x \bar{R}_{xx}-\bar{R}_x R_{xx}+6|R|^2\bar{R}R_x-6|R|^2R \bar{R}_x)\nonumber\\&+i|R|^2,\nonumber\\
\tilde{B}&=16\lambda^4\epsilon R+8\lambda^3\epsilon R_x+4\lambda^2\epsilon(R_{xx}+2|R|^2R)+2\lambda\epsilon(R_{xxx}+6|R|^2R_x)+2i\lambda R\nonumber\\&+\epsilon(R_{xxxx}+8|R|^2R_{xx}+2R^2\bar{R}_{xx}+4|R_x|^2R+6{R}^2_x\bar{R}+6|R|^4R)+iR_x+R.
\end{align}
\end{subequations}

In the above, $\lambda$ belongs to the Lax spectrum of the spectral problem (\ref{i3}) whenever $\xi$ is real.  Since the function $R$ is periodic in space we can represent the solution of the first equation (\ref{i3}) in the form \cite{bdec,dep}
\begin{equation}
\xi(x)=\hat{\xi}(x)e^{i\theta x},
\label{i33}
\end{equation}
where $\hat{\xi}(x)$ is a function periodic in space and $\theta$ is the Floquet parameter ($\theta \in [0,\frac{\pi}{T}]$).  The second equation in (\ref{i3}) is a linear algebraic system of equations.  A non-zero solution to this system of equations can be found by demanding the determinant of the coefficient matrix is zero.  Evaluating the determinant, we find
\begin{align}
\Omega^2+Q(\lambda)=0,
\label{i5}
\end{align}
where 
\begin{align}
Q(\lambda)=4\lambda^4-4b\lambda^2+b^2+2d+\lambda^2p+a,
\label{i6}
\end{align}
with $2d=|R_x|^2+|R|^4-2b|R|^2, a=i(R_x\bar{R}-R\bar{R}_x)+|R|^2$ and $p=8|R|^2-1$. To make $Q(\lambda)$ independent of $x$ we restrict $a=c=p=0$ so that Eq. (\ref{i6}) further reduces to
\begin{align}
Q(\lambda)=4\tilde{Q}(\lambda),
\label{i7}
\end{align}
where 
\begin{align}
\tilde{Q}(\lambda)=\lambda^4-2\tilde{b}\lambda^2+\tilde{b}^2+2\tilde{d},\quad b=2\tilde{b},\quad d=2\tilde{d}.
\label{i8}
\end{align}

The quartic polynomial $\tilde{Q}(\lambda)$ can be rewritten as 
\begin{align}
\tilde{Q}(\lambda)=\lambda^4-\frac{1}{2}(R^2_1+R^2_2)\lambda^2+\frac{1}{16}(R^2_1-R^2_2)^2,
\end{align}
where we have considered $\tilde{b}=\frac{1}{4}(R^2_1+R^2_2)$ and $\tilde{d}=-\frac{1}{8}R^2_1R^2_2$.  With this choice, the quartic polynomial $\tilde{Q}(\lambda)$ can be factorized as
\begin{align}
\tilde{Q}(\lambda)=(\lambda^2-\lambda^2_1)(\lambda^2-\lambda^2_2),
\label{i9}
\end{align}
where $\lambda_1=\pm \frac{R_1+R_2}{2}$ and $\lambda_2=\pm \frac{R_1-R_2}{2}$. In other words, Eq. (\ref{i5}) has now been rewritten in a simplified form as $\Omega^2+4\tilde{Q}(\lambda)=0$.

 To determine the value of $\lambda$ we substitute Eqs. (\ref{i2}) and (\ref{i33}) into the Lax pair Eq. (\ref{e2}) so that the eigenvalue problem can be reformulated in the form
 \begin{equation}
 \begin{pmatrix} \frac{d}{dx}+i\theta & -R\\ -\bar{R} & -(\frac{d}{dx}+i\theta) \end{pmatrix}\begin{pmatrix} \hat{\xi}_1\\ \hat{\xi}_2 \end{pmatrix}=\lambda \begin{pmatrix} \hat{\xi}_1\\ \hat{\xi}_2 \end{pmatrix},
 \label{i99}
 \end{equation}
 where $\theta\in(0,\frac{\pi}{T})$.  Upon solving the eigenvalue Eq. (\ref{i99}) we can compute the Lax spectrum.  We solve Eq. (\ref{i99}) only numerically \cite{chen4,bdec}.

Now, let us add a linear perturbation term $S(x-ct,t)$ to the periodic wave/seed solution (\ref{i1}) in the form
\begin{align}
r(x,t)=[R(x-ct)+S(x-ct,t)]e^{2ibt}.
\label{i10}
\end{align}
Substituting Eq. (\ref{i10}) into (\ref{e1}) and dropping the quadratic terms in $S$, we obtain
\begin{subequations}
\label{subi1}
\begin{align}
\label{i11}
&iS_t-2bS-icS_x+S_{xx}+4|R|^2S+2R^2\bar{S}-iS_x-i\epsilon(S_{xxxxx}+10|R|^2S_{xxx}\nonumber\\
&+10RR_x\bar{S}_{xx}+10R \bar{R}_xS_{xx}+20\bar{R}R_xS_{xx}+10R R_{xx}\bar{S}_x+10R^2_x\bar{S}_x+10R\bar{R}_{xx}S_x\nonumber\\
&+20\bar{R}R_{xx}S_x+20|R_x|^2S_x+30|R|^4S_x+10RR_{xxx}\bar{S}+10\bar{R}R_{xxx}S+10R_x \bar{R}_{xx}S\nonumber\\
&+10\bar{R}_xR_{xx}S+20R_x R_{xx}\bar{S}+60|R|^2RR_x\bar{S}+60|R|^2\bar{R}R_xS)=0, \\\nonumber\\
&-i\bar{S}_t-2b\bar{S}+ic\bar{S}_x+\bar{S}_{xx}+4|R|^2\bar{S}+2\bar{R}^2S+iS_x+i\epsilon(\bar{S}_{xxxxx}+10|R|^2\bar{S}_{xxx}\nonumber\\
&+10\bar{R}\bar{R}_xS_{xx}+10\bar{R}R_x\bar{S}_{xx}+20R\bar{R}_x\bar{S}_{xx}+10\bar{R}\bar{R}_{xx}S_x+10\bar{R}^2_xS_x+10\bar{R}R_{xx}\bar{S}_x\nonumber\\
&+20R\bar{R}_{xx}\bar{S}_x+20|R_x|^2\bar{S}_x+30|R|^4\bar{S}_x+10\bar{R}\bar{R}_{xxx}S+10R\bar{R}_{xxx}\bar{S}+10\bar{R}_x R_{xx}\bar{S}\nonumber\\
&+10R_x\bar{R}_{xx}\bar{S}+20\bar{R}_x \bar{R}_{xx}S+60|R|^2\bar{R}\bar{R}_xS+60|R|^2R\bar{R}_x\bar{S})=0.
\end{align}
\end{subequations}

We choose a separable solution for the functions $S(x,t)$ and $\bar{S}(x,t)$ in the form
\begin{align}
S(x,t)=v_1(x)e^{t\Lambda}, \quad \bar{S}(x,t)=v_2(x)e^{t\Lambda},
\label{i12}
\end{align}
where $\Lambda$ is the spectral parameter and $v=(v_1,v_2)^{T}$ satisfies the spectral stability problem,
\begin{align}
\label{i13}
i\Lambda\sigma_3 v+\begin{pmatrix} m_1 & m_2\\ \bar{m}_2 & \bar{m}_1 \end{pmatrix}v=0, \quad
\sigma_3 = \begin{pmatrix}
1 & 0 \\
0 & -1
\end{pmatrix},
\end{align}
with
\begin{subequations}
	\label{subi2}
\begin{align}
m_1=&-2b-ic\partial_x+\partial_{xx}+4|R|^2-i\partial_x-i\epsilon(\partial_{xxxxx}+10|R|^2\partial_{xxx}+10R \bar{R}_x\partial_{xx}\nonumber\\
&+20\bar{R}R_x\partial_{xx}+10R\bar{R}_{xx}\partial_x+20\bar{R}R_{xx}\partial_x+20|R_x|^2\partial_x+30|R|^4\partial_x\nonumber\\
&+10\bar{R}R_{xxx}+10R_x \bar{R}_{xx}+10\bar{R}_xR_{xx}+60|R|^2\bar{R}R_x), \\ 
m_2=&2R^2-i\epsilon(60|R|^2RR_x+20R_x R_{xx}+10RR_{xxx}+10R^2_x\partial_x+10R R_{xx}\partial_x\nonumber\\
&+10R R_x\partial_{xx}).
\end{align}
\end{subequations}

In the above $\Lambda$ refers the stability spectrum of spectral problem (\ref{i13}) provided $v$ is real.  On the other hand, the eigenvalue $\lambda$ is associated with the Lax spectrum of the spectral problem (\ref{i3}).  The bounded squared eigenfunctions $\xi^2_1$ and $\xi^2_2$ determine the bounded eigenfunctions of Eq. (\ref{i13}), that is, $v_1=\xi^2_1$ and $v_2=-\xi^2_2$  and $\Omega$ determines the eigenvalues through the relation $\Lambda=2\Omega$ (a  rigorous proof for the connection between the squared eigenfunctions and the stability parameter - one may refer \cite{bdec} and references therein).  Substituting the relation $\Lambda=2\Omega$ in the expression $\Omega^2+4\tilde{Q}(\lambda)=0$, we find $\Lambda^2 + 16 \tilde{Q}(\lambda)=0$.  This relation connects the eigenspectrum of the unperturbed part with perturbed part.  As a result, now we are in a position to calculate the instability growth rate through the relation $Re(\Lambda)=\pm Re\left(4i\sqrt{\tilde{Q}(\lambda)}\right)$.  Now, we investigate how the instability growth rate of both $dn$- and $cn-$ periodic waves modify while we change the value of the elliptic modulus parameter $k$.
\begin{figure}[!ht]
	\begin{center}
		\begin{subfigure}{0.47\textwidth}
			\caption{}
			\includegraphics[width=\linewidth]{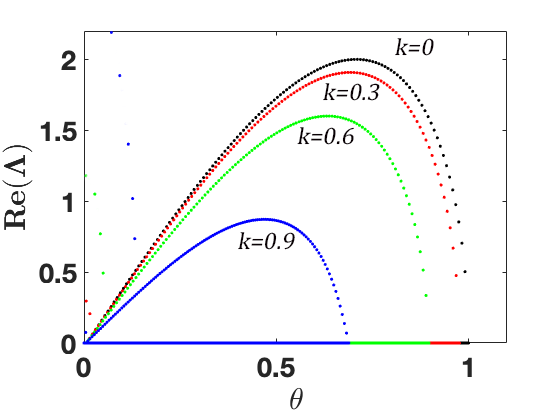}
		\end{subfigure}
		\begin{subfigure}{0.47\textwidth}
			\caption{}
			\includegraphics[width=\linewidth]{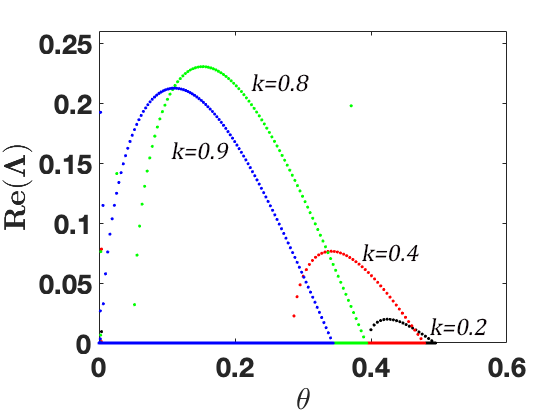}
		\end{subfigure}
	\end{center}
	\vspace{-0.6cm}
	\caption{{ Modulational instability growth rate Re($\Lambda$) versus the Floquet parameter $\theta$ for (a) $dn$-periodic waves and (b) $cn$-periodic waves.}}
	\label{dnnfig5}
\end{figure}

In Fig. \ref{dnnfig5} we show the instability growth rate of both $dn$ and $cn$ periodic waves for four distinct values of the elliptic modulus parameter. In this figure, Re($\Lambda$) is plotted against the Floquet parameter ($\theta\in [0,\frac{\pi}{T}]$).  The maximum growth rate for the $dn$ periodic wave (left) is realized for $k=0$ (constant-amplitude wave). When we increase the $k$ value from $0$ to $1$ the growth rate decreases continuously which can be seen from Fig. \ref{dnnfig5}(a). In the case of $cn$-periodic waves (right), when we vary the $k$ value from a lower to higher ($0$ to $0.8$) value the growth rate gradually increases.   The maximum growth rate is obtained for $k=0.8$. Beyond this value the growth rate decreases which can be observed from Fig.\ref{dnnfig5}(b).  From these observations, we conclude that the maximum growth rate of $cn$-periodic waves is lower than that of $cn$-periodic waves.  

In Sec. 4, we have illustrated the spatio-temporal dynamics of RW solutions (\ref{e39}) in Figs. \ref{dnnfig3} - \ref{dnnfig41} for three different $k$ values where we have observed that the maximum amplitude of RW on the $dn$- periodic waves is higher than that of the RWs which emerge on the $cn$- periodic background. A similar observation has also been obtained while investigating the instability growth rate (see Fig. \ref{dnnfig5}). Thus the emergence of RWs is intimately related with the modulational instability of background wave.
\section{Conclusion}
In this work, we have constructed RW solutions on the $dn$ and $cn$ periodic wave background for the FONLS equation with the help of DT and the method of nonlinearization of Lax pair.  Using the latter method, we have determined the eigenvalues and squared eigenfunctions that correspond to the elliptic travelling wave solutions of the FONLS equation. From the obtained eigenvalues, eigenfunctions and the seed solution choosen, we have created the periodic background solution.  To construct the RWs on top of this periodic wave background we have considered second independent solution (in a non-periodic form) for the spectral problem. We have analyzed the constructed RWs for three different values of the system parameter $(\epsilon)$.  We have noticed that frequency of periodic background wave changes in the $(x-t)$ plane when we increase the value of the system parameter. We have also evaluated the instability rate of $dn$- and $cn$- periodic waves for the FONLS equation through the spectral stability problem. Our investigations demonstrate that when we increase the elliptic modulus value, the amplitude of RWs on the periodic background decreases (increases) for $dn$ ($cn$) wave.  Furthermore, we have noticed that the maximum amplitude of RW on the $cn$-periodic background is lower than that of $dn$-periodic background.  Our results will be useful in understanding some features that are associated with the localized wave phenomena that originate from higher order effects.  

\section*{Acknowledgments}
NS thanks the University for providing University Research Fellowship.  KM wishes to thank the Council of Scientific and Industrial Research, Government of India, for providing the Research Associateship under the Grant No. 03/1397/17/EMR-II. The work of MS forms part of a research project sponsored by National Board for Higher Mathematics, Government of India, under the Grant No. 02011/20/2018NBHM(R.P)/R\&D 24II/15064. 



\begin{thebibliography}{90} 
	\bibitem{ankiew}
	N. Akhmediev and A. Ankiewicz, Solitons: Nonlinear Pulses and Beams (Chapman \& Hall, London, 1997).
	
	\bibitem{obsorne}
	A. Osborne, Nonlinear Ocean Waves and the Inverse Scattering Transform (Academic Press, London, 2010).
	
	\bibitem{dudley}
	J. M. Dudley, G. Genty, A. Mussot, A. Chabchoub and F. Dias, Nat. Rev. Phys. {\bf 1}, 675-689 (2019).
	
	\bibitem{cha}
	A. Chabchoub, N. Hoffmann and N. Akhmediev, Phys. Rev. Lett. {\bf 106}, 204502 (2011).
	
	\bibitem{so}
	D. R. Solli, C. Ropes, P. Kovnath, and B. Jalali, Nature {\bf 450}, 1054-1057 (2007).
	
	\bibitem{kib}
	B. Kibler, J. Fatome, C. Finot, G. Millot, F. Dias, G. Genty, N. Akhmediev and J. M. Dudley, Nat. Phys. {\bf 6}, 790-795 (2010).
	
	\bibitem{mani}
	K. Manikandan, P. Muruganandam, M. Senthilvelan and M. Lakshmanan, Phys. Rev. E.  {\bf 90},  062905 (2014).
	
		\bibitem{blu}
	Y. V. Bludov, V. V. Konotop and  N. Akhmediev, Phys. Rev. A.  {\bf 80},  033610 (2009).
	
	
		\bibitem{nak}
	N. Akhmediev, A. Ankiewicz and M. Taki, Phys. Lett. A. {\bf 373}, 675-678 (2009).	
	
	\bibitem{chen4}
	J. Chen, D. E. Pelinovsky and R. E. White, Physica D {\bf 405}, 132378 (2020).
	
		\bibitem{ye}
	Y. Ye, J. Liu, L. Bu, C. Pan, S. Chen and D. Mihalache, Nonlinear Dyn. {\bf 102}, 1801-1812 (2020).
	
	\bibitem{zakharov}
	V. E. Zakharov and L. A. Ostrovsky, Physica D {\bf 238}, 540-548 (2009).
	
	\bibitem{agaf}
	D. S. Agafontsev and V. E. Zakharov, Nonlinearity {\bf 28},  2791-2821 (2015).
	
	\bibitem{mu}
	G. Mu, Z. Qin and R. Grimshaw, SIAM J. Appl. Math. {\bf 75},  1-20 (2015).
	
	\bibitem{agaf1}
	D. S. Agafontsev and V. E. Zakharov, Nonlinearity {\bf 29}, 3551-3578 (2016).
	
	\bibitem{mani2}
	K. Manikandan, P. Muruganandam, M. Senthilvelan and M. Lakshmanan, Phys. Rev. E.  {\bf 93}, 093202 (2016).
	
	\bibitem{chen5}
	J. Chen, D. E. Pelinovsky and R. E. White, Phys. Rev. E {\bf 100}, 052219 (2019). 
	
	
	\bibitem{kedziora}
	D. J. Kedziora, A. Ankiewicz and N. Akhmediev, Euro. Phys. J. Spec. Topics {\bf 223}, 43-62 (2014).

	\bibitem{chen2}
	J. Chen and D. E. Pelinovsky, Proc. R. Soc. A {\bf 474}, 20170814 (2018). 
	
	\bibitem{chen1}
	J. Chen and D. E. Pelinovsky, Nonlinearity {\bf 31}, 1955 (2018).
	
	\bibitem{chen3}
	J. Chen and D. E. Pelinovsky, J. Nonlinear Sci. {\bf 29}, 2797 (2019). 
	
	\bibitem{sinthu3}
	N. Sinthuja, K. Manikandan and M. Senthilvelan, Phys. Scr. {\bf 96}, 105206 (2021).
	
	\bibitem{li}
	R. Li and X. Geng, Appl. Math. Lett. {\bf 102}, 106147 (2020).
		
	\bibitem{hirota}
	R. Hirota, J. Math. Phys. 14, 805 (1973).
		
	\bibitem{agrawal}
	G. P. Agrawal, Nonlinear Fiber Optics - $6^{th}$ Ed. (Academic Press, London, 2019).
	
	\bibitem{akmv1}
	A. Ankiewicz, J. M. Soto-Crespo and N. Akhmediev, Phys. Rev. E 81, 046602 (2010).
		
	\bibitem{akmv2}
	A. Ankiewicz, J. M. Soto-Crespo, M. A. Chowdury and N. Akhmediev, J. Opt. Soc. Am. B {\bf 30}, 87-94 (2013).

	\bibitem{porse}
M. Lakshmanan, K. Porsezian and M. Daniel, Phys. Lett. A {\bf 133}, 483 (1988).

	\bibitem{porse1}
K. Porsezian, M. Daniel and M. Lakshmanan, J. Math. Phys. {\bf 33}, 1807 (1992).

	\bibitem{wzzhao}
W. Z. Zhao, Y. Q. Bai and K. Wu, Phys. Lett. A {\bf 32}, 64-68 (2006).

	\bibitem{radha}
R. Radha and V. R. Kumar, Z. Naturforsch. A {\bf 62}, 381–386 (2007).

   \bibitem{akmv3}
   A. Chowdury, D. J. Kedziora, A. Ankiewicz and N. Akhmediev, Phys. Rev. E {\bf 90}, 032922 (2014).

	\bibitem{akmv4}
   A. Ankiewicz, D. J. Kedziora, A. Chowdury, U. Bandelow and N. Akhmediev, Phys. Rev. E {\bf 93}, 012206 (2016).
   
   \bibitem{sun}
    W. R. Sun, B. Tian, H. L. Zhen and Y. Sun, Nonlinear Dyn. {\bf 81}, 725-732 (2015).
    
    \bibitem{song}
	N. Song, H. Xue and X. Zhao, IEEE Access {\bf 8}, 9610-9618 (2020).
	
	 \bibitem{peng}
	W. Q. Peng, S. F. Tian, X. B. Wang and T. T. Zhang, Wave Motion {\bf 93}, 102454 (2020).
	
	\bibitem{zhang}
	H. Q. Zhang and F. Chen, Chaos {\bf 31}, 023129 (2021).
	
	\bibitem{zhagao}
    H. Q. Zhang, X. Gao, Z. J. Pei and F. Chen, Appl. Math. Lett. {\bf 107}, 106464 (2020).
   
	\bibitem{sin}
		N. Sinthuja, K. Manikandan and M. Senthilvelan, Euro. Phys. J.  Plus  {\bf 136}, 305 (2021).
		
	\bibitem{wang}
	P. Wang, Euro. Phys. J. D {\bf 68} 181 (2014).
	
	\bibitem{feng}
	L. L. Feng, S. F. Tian and T. T. Zhang, Rocky Mountain J. Math. {\bf 49}, 29-45 (2019).
  
    \bibitem{backus}
    S. Backus, C. G. Durfee, G. Mourou, H.C. Kapten and M.M. Murnane, Opt. Lett. {\bf 22}, 1256 (1997).
    
    \bibitem{christov}
    I. P. Christov, Phys. Rev. A {\bf 60}, 3244 (1999).
    
    \bibitem{henkel}
    J. Henkel, T. Witting, D. Fabris and J.P. Marangos, Phys. Rev. A {\bf 87}, 043818 (2013).
    
    \bibitem{wang2}
    Q. M. Wang, Y. T. Gao, C. Q. Su, Y. J. Shen, Y. J. Feng and L.Xue, Z. Naturforsch. A {\bf 70}, 365-374 (2015).
		
\bibitem{darboux}	
	V. B. Matveev and M. A. Salle, Darboux Transformation and Solitons. (Springer, Berlin (1991)).

    \bibitem{zhou}
    R. G. Zhou, J. Math. Phys. {\bf 48}, 013510 (2007).

    \bibitem{zhou1}
    R. G. Zhou, Stud. Appl. Math. {\bf 123}, 311 (2009).
    
    \bibitem{hase}
    A. Hasegawa, Modulational Instability. In: Optical Solitons in Fibers. (Springer, Berlin, Heidelberg, 1990).

	\bibitem{bdec}
	B. Deconinck and B. Segal, Physica D {\bf 346}, 1 (2017).
	
		\bibitem{gxu}
	G. Xu, A. Chabchoub, D. E. Pelinovsky and B. Kibler, Phys. Rev. Research {\bf 2}, 033528 (2020).
	
		\bibitem{dep}
	 D. E. Pelinovsky, Front. Phys. {\bf 9}, 599146 (2021).

\end{thebibliography}
\end{document}